\documentclass{cimart}

\usepackage{enumitem}

\usepackage{float}

\DeclareMathOperator{\Novel}{Novel}
\DeclareMathOperator{\Dyck}{Dyck}
\DeclareMathOperator{\FacEq}{FacEq}
\DeclareMathOperator{\Bal}{Bal}
\DeclareMathOperator{\Nest}{Nest}
\DeclareMathOperator{\per}{per}
\DeclareMathOperator{\ce}{ce}

\newcommand{\seqnum}[1]{\href{https://oeis.org/#1}{\rm \underline{#1}}}
\def\modd#1 #2{#1\ \mbox{\rm (mod}\ #2\mbox{\rm )}}
\def\andd{\, \wedge \,}

\newenvironment{smallarray}[1]
{\null\,\vcenter\bgroup\scriptsize
	\arraycolsep=.13885em
	\hbox\bgroup$\array{@{}#1@{}}}
{\endarray$\egroup\egroup\,\null}

\title{Dyck words, pattern avoidance, and automatic sequences}

\author{Lucas Mol, Narad Rampersad and Jeffrey Shallit}

\authorinfo[L. Mol]
{Department of Mathematics and Statistics,
	Thompson Rivers University, Canada}
{lmol@tru.ca}

\authorinfo[N. Rampersad]
{Department of Mathematics and Statistics, University of Winnipeg, Canada}
{n.rampersad@uwinnipeg.ca}

\authorinfo[J. Shallit]
{School of Computer Science, University of Waterloo, Canada}
{shallit@uwaterloo.ca}

\abstract{
	We study various aspects of Dyck words appearing in binary sequences, where $0$ is treated as a left parenthesis and $1$ as a right parenthesis.  We show that binary words that are $7/3$-power-free have bounded nesting level, but this no longer holds for larger repetition exponents.   We give an explicit characterization of the factors of the Thue-Morse word that are Dyck, and show how to count them.  We also prove tight upper and lower bounds on $f(n)$, the number of Dyck factors of Thue-Morse of length $2n$.
}



\VOLUME{33}
\YEAR{2025}
\ISSUE{2}
\NUMBER{5}
\DOI{https://doi.org/10.46298/cm.12695}

\begin{document}
	
	\section{Introduction}
	We define $\Sigma_k := \{ 0,1,\ldots, k-1 \}$. 
	Suppose $x \in \Sigma_2^*$; that is,
	suppose $x$ is a finite binary word.  We say it is a {\it Dyck word} 
	if, considering $0$ as a left parenthesis and $1$ as a right parenthesis, the word represents a string of balanced parentheses \cite{Chomsky&Schutzenberger:1963}.  For example, $010011$ is Dyck,
	while $0110$ is not.  Formally, $x$ is Dyck if $x$ is empty, or there
	are Dyck words $y, z$ such that either $x = 0y1$ or
	$x = yz$.   The set of all Dyck words forms the {\it Dyck language}.
	
	In this paper we are concerned with the properties of factors of infinite binary words that are Dyck words.  
	
	If $x$ is a Dyck word, we may talk about its {\it nesting level\/} $N(x)$, which is the deepest level of parenthesis nesting in the string it represents.   Formally,  we have that $N(\epsilon) = 0$,
	$N(0y1) = N(y) + 1$, and
	$N(yz) = \max(N(y), N(z))$ if
	$y, z$ are Dyck words.   The Dyck property and nesting level are intimately connected with {\it balance}, 
	which is a function defined by $B(x) = |x|_0 - |x|_1$, the excess of $0$'s over $1$'s in $x$.  It is easy to see that a word is Dyck if and only if $B(x) = 0$ and
	$B(x') \geq 0$ for every prefix $x'$ of $x$.  Furthermore, the nesting level of a Dyck word $x$ is the 
	maximum of $B(x')$ over all prefixes $x'$ of $x$.
	
	In this paper we will also be concerned with pattern avoidance, particularly avoidance of powers.   We say a finite word $w = w[1..n]$ has period $p \geq 1$ if $w[i] = w[i+p]$ for all indices $i$ with $1 \leq i \leq n-p$.   The smallest period of $w$ is called {\it the\/} period, and is denoted $\per(w)$.
	The {\it exponent\/} of a finite word $w$ is defined to be $\exp(w) := |w|/\per(w)$.   A word with exponent $\alpha$ is said to be an
	$\alpha$-power.  For example,
	$\exp({\tt alfalfa}) = 7/3$ and so
	{\tt alfalfa} is a $7/3$-power.  If a word contains no powers $\geq \alpha$, then we say it is {\it $\alpha$-power-free}.   If it contains no powers $> \alpha$, then we say it is
	{\it $\alpha^+$-power-free}.  If $w$ is a finite or infinite word, its {\it critical exponent} is defined to be
	$\ce(w) := \sup \{ \exp(x)\colon \, x \text{ is a finite nonempty factor of } w \}$.    A \emph{square} is a word of the form $xx$, where $x$ is a nonempty word.  An {\it overlap\/} is a word of the form
	$axaxa$, where $a$ is a single letter and $x$ is a possibly empty word.
	
	Some of our work is carried out using the {\tt Walnut} theorem prover, which can rigorously prove many results about automatic sequences.   See \cite{Mousavi:2016,Shallit:2022} for more details.
	{\tt Walnut} is free software that can be downloaded at \\
	\centerline{\url{https://cs.uwaterloo.ca/~shallit/walnut.html} \ .}
	
	A preliminary version of this paper appeared
	previously \cite{Mol}.
	
	\section{Repetitions and Dyck words}
	\label{Sec:OverlapFree} 
	
	\begin{theorem}\label{nesting}
		If a binary word is $7/3$-power-free and Dyck, then its nesting level is at most $3$.
	\end{theorem}
	
	\begin{proof}
		The $7/3$-power-free Dyck words of nesting level $1$ are $01$ and $0101$.
		The set of $7/3$-power-free Dyck words of nesting level $2$ is therefore a subset of
		$\{01,0011,001011\}^*$.  Let $x$ be a $7/3$-power-free Dyck word of nesting level $3$.
		Suppose that $x=0y1$, where $y$ has nesting level $2$.  Then, to avoid the cubes $000$
		and $111$, the word $y$ must begin with $01$ and end with $01$.  Furthermore,
		since $y$ has nesting level $2$ it must contain one of $0011$ or $001011$.
		Write $x=001y'011$.  The word $y'$ cannot begin or end with $01$, since that would
		imply that $x$ contains one of the $5/2$-powers $01010$ or $10101$.  Thus $y'$ begins with $001$
		and ends with $011$, which means $x$ begins with $001001$ and ends with $011011$.
		Consequently $x$ cannot be extended to the left or to the right without creating
		a cube or $7/3$-power.  Furthermore, this implies that a $7/3$-power-free Dyck word of nesting
		level $3$ cannot be written as a concatenation of two non-empty Dyck words, nor can it be
		extended to a $7/3$-power-free Dyck word of nesting level $4$.
	\end{proof}
	
	\begin{theorem}\label{Thm:OverlapFreeChar}
		Define $h(0) = 01$, 
		$h(1) = 0011$, and $h(2) = 001011$.  A binary word $w$ is an overlap-free Dyck word if and only if either
		\begin{enumerate}[label=\normalfont(\roman*)]
			\item $w=h(x)$, where $x\in\Sigma_3^*$ contains no square as a proper factor and contains no $212$ or $20102$; or
			\item $w=0h(x)1$, where $x\in \Sigma_3^*$ is square-free, begins with $01$ and ends with $10$, and contains no $212$ or $20102$.
		\end{enumerate}
	\end{theorem}
	
	\begin{proof}
		Let $w$ be an overlap-free Dyck word.  By Theorem~\ref{nesting}, we have $N(w) \leq 3$.  Suppose $N(w) \leq 2$.
		Then $w \in \{01,0011,001011\}^*$ by the proof of Theorem~\ref{nesting}.  So, we have $w=h(x)$ for some
		$x \in \Sigma_3^*$.  If $N(w)=3$, then by the proof of Theorem~\ref{nesting}, we have $w=0h(x)1$.
		If $x$ contains a square $yy$ as a proper factor, then certainly $w$ contains one of the overlaps
		$1h(y)h(y)$ or $h(y)h(y)0$.  Furthermore, if $x$ contains $212$, then $w$ contains the overlap
		$011001100$ and if $x$ contains $20102$, then $w$ contains the overlap $1101001101001$.
		Finally, if $w=0h(x)1$, then $x$ must begin and end with $0$ and contain at least one $1$ or $2$.
		If $x$ begins with $02$, then $w$ contains the overlap $0010010$, and if $x$ ends with $20$, then $w$ contains
		the overlap $1011011$.  Thus, $x$ begins with $01$ and ends with $10$.
		
		For the other direction, let $x\in \Sigma_3^*$ be a squarefree word that contains no $212$ or $20102$.  First consider the word $h(x)$, which is clearly a Dyck word.  We now show that $h(x)$ is overlap-free.  We verify by computer that if $|x|\leq 10$, then $h(x)$ is overlap-free.  So, we may assume that $|x|\geq 11$.  Suppose towards a contradiction that $h(x)$ contains an overlap $z$.  Assume that $z=0y0y0$; the case $z=1y1y1$ is similar, and the proof is omitted.  We consider several cases depending on the prefix of $y$.
		
		If $y$ starts with $0$, then $h^{-1}(z0^{-1})=h^{-1}(0y0y)$ is a square that appears as a proper factor of $x$.
		
		If $y$ starts with $100$, write $y=100y'$, so that $z=0100y'0100y'0$.  In this case, $h^{-1}(z0^{-1})=h^{-1}(0100y'0100y')$ is a square that appears as a proper factor of $x$.
		
		If $y$ starts with $101$, write $y=101y'$, so that $z=0101y'0101y'0$.  Note that $00$ is not a factor of $x$, so any occurrence of $0101$ in $z$ is as a factor of
		$h(2) = 001011$.  Consequently, the word $h^{-1}(0z0^{-1})=h^{-1}(00101y'00101y')$ is a square that appears as a proper factor of~$x$.
		
		Finally, if $y$ starts with $11$, then write $y=11y'$, so that $z=011y'011y'0$.  Then $z$ is a factor of $h(ax'bx'c)$, where $a,b,c \in \{1,2\}$, and the value of $b$
		is determined by the suffix of $y'$: if $y'$ ends with $001$ then $b=2$ and if $y'$ ends with $0$ then $b=1$.
		Clearly, we have $a \neq b$ and $b \neq c$, since otherwise $x$ contains a square as a proper factor.  However, if $b=2$ then $y'$ ends with $001$, which
		implies $c=2$, a contradiction.  So, we have $b=1$, and further, since $a \neq b$ and $b \neq c$, we have $a=c=2$.  We therefore have a
		factor $2x'1x'2$ of $x$.  Now $x'$ can neither begin nor end with $2$ or $1$, so we have $2x'1x'2 = 20x''010x''02$.  Similarly, the word
		$x''$ can neither begin nor end with $0$ or $1$, so we have $20x''010x''02 = 202x'''20102x'''202$, whence $x$ contains the forbidden
		factor $20102$, a contradiction.
		
		Thus, we conclude that $h(x)$ is an overlap-free Dyck word.  Finally, assume that $x$ begins with $01$ and ends with $10$, and consider the word $0h(x)1$.  Again, it is clear that $0h(x)1$ is a Dyck word, and we have already shown that the word $h(x)$ is overlap-free.
		Now $0h(x)1$ begins with $0010011$ and ends with $0011011$.  Note that the only occurrences of
		$00100$ and $11011$ as factors of $0h(x)1$ are as a prefix and a suffix, respectively.  It follows that if $0h(x)1$ contains
		an overlap, then this overlap has period at most $4$ and occurs as either a prefix or a suffix of $0h(x)1$.  However,
		one easily verifies that no such overlap exists.  This completes the proof.
	\end{proof}
	
	\begin{corollary}\label{Cor:ofree_nesting}
		There are arbitrarily long overlap-free Dyck words of nesting levels $2$ and $3$.
	\end{corollary}
	
	\begin{proof}
		Consider the well-known word $\bf s$, which is the infinite fixed point, starting with $0$, of the morphism defined by $0\mapsto 012$, $1\mapsto 02$, $2\mapsto 1$.  Thue~\cite{Thue:1912} proved that $\bf s$ is squarefree and contains no $010$ or $212$; this is also easy to verify with \texttt{Walnut} (cf.~\cite{Shallit:2022}).  Let $x$ be a prefix of $\bf s$ that ends in $10$.  Since the factor $10$ appears infinitely many times in $\bf s$, there are arbitrarily long such words $x$.  So, $x$ is squarefree, contains no $212$ or $20102$, begins in $01$, and ends in $10$.  By Theorem~\ref{Thm:OverlapFreeChar}, the words $h(x)$ and $0h(x)1$ are overlap-free Dyck words.  It is easy to see that $h(x)$ has nesting level $2$, and $0h(x)1$ has nesting level $3$,  which completes the proof.
	\end{proof}
	
	The third author and Zavyalov \cite[Theorem~2]{Shallit&Zavyalov:2023} have given an alternative proof of
	Corollary~\ref{Cor:ofree_nesting} (for nesting level $3$).  Their construction uses an implementation of transducers in Walnut to compute and output the nesting level of a word $x$
	if it is $\leq 3$ and output $4$ otherwise.
	
	% ALTERNATE PROOF
	% \begin{proof} (Sketch; details must be filled in.)
		% Let $\mu$ be the Thue-Morse morphism, defined by $\mu(0) = 01$ and $\mu(1) = 10$.
		% Define a sequence of words
		% as follows:  
		% \begin{align*}
			%     A_1 &= 1 \mu^2(00)1 \\
			%     A_{n+1} &= 1 \mu(0 \mu(A_n)1)1, \quad \text{for $n \geq 1$}.
			% \end{align*}
		% Then $00\mu(A_n) 11$ is an overlap-free Dyck word of nesting level $3$.
		% \end{proof}
	
	% \begin{Question}
		% What is the relationship between nesting level of Dyck words and critical exponent?   I suspect that for all nesting levels, there is a finite binary word avoiding $(7/3)^+$ powers with that nesting level.  For example
		% $$ 001001011001001100101100100110110010110011011001011011$$ is a Dyck word of nesting level 5 avoiding $(7/3)^+$-powers.
		% \end{Question}

	Theorem~\ref{nesting} says that every $7/3$-power-free Dyck word has nesting level at most $3$.  We will see that this result is best possible with respect to the exponent $7/3$; in fact, there are $7/3^+$-power-free Dyck words of every nesting level.  Before we proceed with the construction of such words, we provide a very simple construction of cube-free Dyck words of every nesting level, which serves as a preview of the main ideas in the more complicated construction of $7/3^+$-power-free Dyck words of every nesting level.
	
	\begin{lemma}\label{Lemma:BinaryMorphismTrick}
		Let $u$ and $v$ be Dyck words, and let $f:\Sigma_2^*\rightarrow \Sigma_2^*$ be the morphism defined by $f(0)=0u$ and $f(1)=v1$.
		If $w$ is a nonempty Dyck word, then $f(w)$ is a Dyck word, and $N(f(w))=N(w)+\max(N(u),N(v))$.
	\end{lemma}
	
	\begin{proof}
		The proof is by induction on $|w|$.  In the base case, if $w=01$, then $f(w)=0uv1$, and $N(f(w))=1+\max(N(u),N(v))=N(w)+\max(N(u),N(v))$.
		
		Now suppose that $|w|=n$ for some $n>2$, and that the statement holds for all nonempty Dyck words of length less than $n$.  We have two cases.

		\noindent
		\textbf{Case 1:} We have $w=0y1$ for some nonempty Dyck word $y$.

		\noindent
		By the induction hypothesis, the word $f(y)$ is a Dyck word with \[N(f(y))=N(y)+\max(N(u),N(v)).\]  So $f(w)=0uf(y)v1$ is a Dyck word with 
		\begin{align*}
			N(f(w))&=1+\max(N(u),N(f(y)),N(v))\\
			&=1+N(y)+\max(N(u),N(v))\\
			&=N(w)+\max(N(u),N(v)).
		\end{align*}

		\noindent
		\textbf{Case 2:} We have $w=yz$ for some nonempty Dyck words $y,z$. 
		
		\noindent
		By the induction hypothesis, the word $f(y)$ is a Dyck word with $$N(f(y))=N(y)+\max(N(u),N(v)),$$ and $f(z)$ is a Dyck word with $N(f(z))=N(z)+\max(N(u),N(v))$.  Therefore, the word $f(w)=f(y)f(z)$ is a Dyck word with 
		\begin{align*}
			N(f(w))&=\max(N(f(y)),N(f(z)))\\
			&=\max(N(y),N(z))+\max(N(u),N(v))\\
			&=N(w)+\max(N(u),N(v)). \qedhere 
		\end{align*}
	\end{proof}
	
	\begin{corollary}
		There is a cube-free Dyck word of every nesting level.
	\end{corollary}
	
	\begin{proof}
		Let $f:\Sigma_2^*\rightarrow\Sigma_2^*$ be the morphism defined by $f(0)=001$ and $f(1)=011$.  Note that $f(0)=0u$ and $f(1)=u1$, where $u=01$ is a Dyck word with $N(u)=1$.  It is also well-known that the morphism $f$ is cube-free; for example, this follows easily from a criterion of Ker\"anen~\cite{Keranen1984}, which states that to confirm that a uniform binary morphism is cube-free, it suffices to check that the images of all words of length at most 4 are cube-free.  Thus, by a straightforward induction using Lemma~\ref{Lemma:BinaryMorphismTrick}, we see that $w_t=f^t(01)$ is a cube-free Dyck word with $N(w_t)=t+1$.
	\end{proof}
	
	We now define the specific morphisms involved in our construction of $7/3^+$-power-free Dyck words of arbitrarily large nesting level.  Let $g:\Sigma_3^*\rightarrow\Sigma_3^*$ be the $6$-uniform morphism defined by 
	\begin{align*}
		g(0)&=022012,\\
		g(1)&=022112, \text{ and}\\
		g(2)&=202101.
	\end{align*}  
	Let $f:\Sigma_3^*\rightarrow \Sigma_2^*$ be the $38$-uniform morphism defined by 
	\begin{align*}
		f(0)&=00100110100110010110010011001011001101,\\
		f(1)&=00101100110100110110011010010110011011, \text{ and}\\
		f(2)&=00101101001101001011001101001011010011.
	\end{align*}
	We will show that for every $t\geq 0$, the word $f(g^t(2))$ is a $7/3^+$-power-free Dyck word of nesting level $2t+2$.  The letters $f$ and $g$ denote these specific morphisms throughout the remainder of this section.
	
	Over the ternary alphabet $\Sigma_3$, we think of the letter $0$ as a left parenthesis, the letter $1$ as a right parenthesis, and the letter $2$ as a Dyck word.  So we will be particularly interested in the ternary words for which the removal of every occurrence of the letter $2$ leaves a Dyck word, and we call these \emph{ternary Dyck words}.
	
	\begin{definition}
		Let $\beta:\Sigma_3^*\rightarrow \Sigma_2^*$ be defined by $\beta(0)=0$, $\beta(1)=1$, and $\beta(2)=\varepsilon$, and let $w\in\Sigma_3^*$.  If $\beta(w)$ is a Dyck word, then we say that $w$ is a \emph{ternary Dyck word}.  In this case, the \emph{nesting level} of $w$, denoted $N(w)$, is defined by $N(w)=N(\beta(w))$.
	\end{definition}
	
	\begin{lemma}\label{Lemma:g}
		Let $w\in \Sigma_3^*$.  If $w$ is a nonempty ternary Dyck word, then $g(w)$ is a ternary Dyck word with $N(g(w))=N(w)+1$.
	\end{lemma}
	
	\begin{proof}
		Throughout this proof, we let $u=01$, a Dyck word with nesting level $1$.  Note that $\beta(g(0))=001=0u$, $\beta(g(1))=011=u1$, and $\beta(g(2))=0101=u^2$.
		
		The proof is by induction on $|\beta(w)|$.  We have two base cases.  If $\beta(w)=\varepsilon$, then $w=2^i$ for some $i\geq 1$, and $N(w)=0$.  We have $\beta(g(w))=u^{2i}$,
		so we see that $g(w)$ is a ternary Dyck word with $N(g(w))=1=N(w)+1$.  If $\beta(w)=01$, then $w=2^i02^j12^k$ for some $i,j,k\geq 0$, and $N(w)=1$.  We have 
		\[
		\beta(g(w))=u^{2i}(0u) u^{2j}(u1)u^{2k}=u^{2i}0u^{2j+2}1u^{2k},
		\]
		so we see that $g(w)$ is a ternary Dyck word with $N(g(w))=2=N(w)+1$, as desired.
		
		Now suppose that $|\beta(w)|=n$ for some $n>2$, and that the statement holds for all ternary Dyck words $w'$ with $|\beta(w')|<n$.  We have two cases.

		\noindent
		\textbf{Case 1:} We have $\beta(w)=0y1$ for some nonempty Dyck word $y$.

		\noindent
		In this case we may write $w=2^i0w'12^j$ for some $i,j\geq 0$, so that $\beta(w')=y$.  By the induction hypothesis, the word $g(w')$ is a ternary Dyck word with $N(g(w'))=N(w')+1$.  It follows that $\beta(g(w))=u^{2i}0u\beta(g(w'))u1u^{2j}$ is a Dyck word, so $g(w)$ is a ternary Dyck word, and 
		\begin{align*}
			N(g(w))&=1+N(g(w'))\\
			&=1+N(w')+1\\
			&=N(w)+1.
		\end{align*}
		
		\noindent
		\textbf{Case 2:} We have $\beta(w)=y_1y_2$ for some nonempty Dyck words $y_1,y_2$.

		\noindent
		Write $w=w_1w_2$ for some $w_1,w_2\in \Sigma_3^*$ such that $\beta(w_1)=y_1$, and $\beta(w_2)=y_2$. By the induction hypothesis, the words $g(w_1)$ and $g(w_2)$ are ternary Dyck words with 
		\[N(g(w_1))=N(w_1)+1 \text{, and } N(g(w_2))=N(w_2)+1.\]
		Therefore, the word $g(w)=g(w_1)g(w_2)$ is a ternary Dyck word with 
		\begin{align*}
			N(g(w))&=\max\left(N(g(w_1)),N(g(w_2))\right)\\
			&=\max(N(w_1)+1,N(w_2)+1)\\
			&=\max(N(w_1),N(w_2))+1\\
			&=N(w)+1.      \qedhere  
		\end{align*}
	\end{proof}
	
	\begin{lemma}\label{Lemma:f}
		Let $w\in\Sigma_3^*$.  If $w$ is a nonempty ternary Dyck word, then $f(w)$ is a Dyck word with $N(f(w))=2N(w)+2$.
	\end{lemma}
	
	\begin{proof}
		Note that $f(0)=0u_10u_2$, $f(1)=u_31u_41$, and $f(2)=v$, where $u_1$, $u_2$, $u_3$, and $u_4$ are Dyck words of nesting level $2$ and length $18$, and $v$ is a Dyck word of nesting level $2$ and length $38$.  
		
		The proof is by induction on $|\beta(w)|$.  We have two base cases.  If $\beta(w)=\varepsilon$, then $w=2^i$ for some $i\geq 1$, and $N(w)=0$.  We have $f(w)=v$,
		so we see that $f(w)$ is a Dyck word with $N(f(w))=2=2N(w)+2$.  If $\beta(w)=01$, then $w=2^i02^j12^k$ for some $i,j,k\geq 0$, and $N(w)=1$.  We have 
		\[
		f(w)=v^i 0u_10u_2 v^j u_31u_41 v^k,
		\]
		so we see that $f(w)$ is a Dyck word with $N(f(w))=4=2N(w)+2$.
		
		Now suppose that $|\beta(w)|=n$ for some $n>2$, and that the statement holds for all ternary Dyck words $w'$ with $|\beta(w')|<n$.  We have two cases.

		\noindent
		\textbf{Case 1:} We have $\beta(w)=0y1$ for some nonempty Dyck word $y$.

		\noindent
		In this case we may write $w=2^i0w'12^j$ for some $i,j\geq 0$, so that $\beta(w')=y$.  By the induction hypothesis, the word $f(w')$ is a Dyck word with $N(f(w'))=2N(w')+2$.  It follows that $f(w)=v^i0u_10u_2f(w')u_31u_41v^j$ is a Dyck word with
		\begin{align*}
			N(f(w))&=2+N(f(w'))\\
			&=2+2N(w')+2\\
			&=2N(w)+2.
		\end{align*}
		
		\noindent
		\textbf{Case 2:} We have $\beta(w)=y_1y_2$ for some nonempty Dyck words $y_1,y_2$.

		\noindent
		Write $w=w_1w_2$ for some $w_1,w_2\in \Sigma_3^*$ such that $\beta(w_1)=y_1$, and $\beta(w_2)=y_2$. By the induction hypothesis, the words $f(w_1)$ and $f(w_2)$ are Dyck words with 
		\[N(f(w_1))=2N(w_1)+2 \text{, and }N(f(w_2))=N(w_2)+1.\]  
		Therefore, the word $f(w)=f(w_1)f(w_2)$ is a Dyck word with 
		\begin{align*}
			N(f(w))&=\max\left(N(f(w_1)),N(f(w_2))\right)\\
			&=\max(2N(w_1)+2,2N(w_2)+2)\\
			&=2\max(N(w_1),N(w_2))+2\\
			&=2N(w)+2.    \qedhere
		\end{align*}
	\end{proof}
	
	\begin{theorem}
		There are $7/3^+$-power-free Dyck words of every nesting level.
	\end{theorem}
	
	\begin{proof}
		Let $t\geq 0$.  We claim that the word $f(g^t(2))$ is a $7/3^+$-free Dyck word of nesting level $2t+2$.  Since $2$ is a ternary Dyck word with nesting level $0$, by Lemma~\ref{Lemma:g}, and a straightforward induction, the word $g^t(2)$ is a ternary Dyck word with nesting level $t$.  Thus, by Lemma~\ref{Lemma:f}, the word $f(g^t(2))$ is a Dyck word with nesting level $2t+2$.
		
		It remains only to show that $f(g^t(2))$ is $7/3^+$-power-free.  We use the \texttt{Walnut} theorem-prover to show that $f(g^\omega(0))$ is $7/3^+$-power-free, which is equivalent.   One only need type in the following commands:
		\begin{verbatim}
			morphism f 
			"0->00100110100110010110010011001011001101 
			1->00101100110100110110011010010110011011 
			2->00101101001101001011001101001011010011":
			
			morphism g "0->022012 1->022112 2->202101":
			
			promote GG g:
			image DFG f GG:
			
			eval DFGtest "?msd_6 Ei,n (n>=1) & At (3*t<=4*n) => 
			DFG[i+t]=DFG[i+t+n]":
		\end{verbatim}
		and {\tt Walnut} returns {\tt FALSE}.   Here the first two {\tt morphism} commands define $f$ and $g$, and the next two commands create a DFAO for $f(g^\omega(0))$.  Finally, the last command asserts the existence of a $7/3^+$ power in $f(g^\omega(0))$.
		
		This was a large computation in {\tt Walnut},  requiring 130 GB of memory and 20321 seconds of CPU time.
	\end{proof}
	
	\begin{remark}
		An alternative method of proof is to first use {\tt Walnut} to show that the word $g^\omega(0)$ is overlap-free, and then apply an extended version~\cite[Lemma 23]{MolRampersadShallit2020} of a well-known result of Ochem~\cite[Lemma 2.1]{Ochem2006} to show that $f(g^\omega(0))$ is $7/3^+$-power-free.
	\end{remark}

	\section{Dyck factors of Thue-Morse}
	
	In this section we give a characterization of those
	factors of $\bf t$, the Thue-Morse sequence, that are
	Dyck.
	
	Let $g:\Sigma_3^* \to \Sigma_2^*$ be the morphism defined by $g(0)=011$, $g(1)=01$, and $g(2)=0$ and let
	$f:\Sigma_3^* \to \Sigma_3^*$ be the morphism defined by $f(0)=012$, $f(1)=02$, and $f(2)=1$.  Define ${\bf s}=f^\omega(0)$.  It is
	well-known (see \cite[Proposition~2.3.2]{Lothaire:1997}) that $g({\bf s})={\bf t}$.  Recall the morphism $h:\Sigma_2^* \to \Sigma_2^*$ defined earlier
	by $h(0) = 01$, $h(1) = 0011$, and $h(2) = 001011$.
	
	\begin{theorem}\label{tm_dyck}
		The Dyck factors of the Thue-Morse word are exactly the words $h(x)$ where $x$ is a factor of ${\bf s}$.
		\label{dycktm}
	\end{theorem}
	
	\begin{proof}
		By considering the return words of $11$ in $\bf t$ (here what we mean are all factors
		$r$ of $\bf t$ that have exactly one occurrence of $11$, as a suffix, and
		always occur in $\bf t$ either as a prefix of $\bf t$ or following an occurrence of $11$;
		see \cite{Balkova&Pelantova&Steiner:2006}) we see that $\bf t$
		begins with $011$ followed by a concatenation of the four words
		$$0011,\quad 010011,\quad 001011,\quad 01001011.$$
		These are all Dyck words, as shown by the bracketings
		$$(0(01)1),\quad (01)(0(01)1),\quad (0(01)(01)1),\quad (01)(0(01)(01)1).$$
		Furthermore, these words must have the above bracketings when they occur as factors
		of any larger Dyck word in $\bf t$.  It follows that ${\bf t}=011{\bf t}'$, where
		${\bf t}'$ is a concatenation of the three Dyck words $h(0) = 01$, $h(1) = 0011$,
		and $h(2) = 001011$.
		
		To complete the proof, it suffices to show that $h({\bf s}) = (011)^{-1}{\bf t} = (011)^{-1}g({\bf s})$.
		We have
		\begin{align*}
			h(f(0)) &= h(012) = g(120210) = g(0^{-1}f^2(0)0)\\
			h(f(1)) &= h(02) = g(1210) = g(0^{-1}f^2(1)0)\\
			h(f(2)) &= h(1) = g(20) = g(0^{-1}f^2(2)0),
		\end{align*}
		so
		$$h({\bf s}) = h(f({\bf s})) = g(0^{-1}f^2({\bf s})) = g(0^{-1}{\bf s}) = (011)^{-1}g({\bf s}),$$
		as required.
	\end{proof}
	
	\section{Dyck factors of some automatic sequences}
	
	In this section we are concerned with Dyck factors of automatic sequences.  Recall that a sequence over a finite alphabet
	$(s(n))_{n \geq 0}$ is {\it $k$-automatic\/} if there exists
	a DFAO (deterministic finite automaton with output) that,
	on input $n$ expressed in base $k$, reaches a state
	with output $s(n)$.
	
	Since the Dyck language is not a member of the FO[+]-definable languages \cite{Choffrut&Malcher&Mereghetti&Palano:2012}, this means that ``automatic'' methods (like that implemented in the {\tt Walnut} system; see \cite{Mousavi:2016,Shallit:2022}) cannot always directly handle such words.
	However, in this section we show that if a $k$-automatic sequence also
	has a certain special property, then the number of Dyck factors
	of length $n$ occurring in it is a $k$-regular sequence.  
	
	To explain the special property, we need the notion of synchronized
	sequence \cite{Shallit:2021h}.   We say a sequence $(v(n))_{n \geq 0}$ is {\it synchronized\/}
	if there is a finite automaton accepting, in parallel, the base-$k$ representations of $n$ and $v(n)$.  Here the shorter representation is padded with leading zeros, if necessary.
	
	Now suppose ${\bf s} = (s(n))_{n \geq 0}$ is a $k$-automatic sequence taking values in $\Sigma_2$ and define the running sum
	sequence $v(n) = \sum_{0 \leq i < n} s(i)$.
	If ${\bf v} = (v(n))_{n \geq 0}$ is synchronized, we say that $\bf s$
	is {\it running-sum synchronized}.  For example, any fixed point of a $k$-uniform binary
	morphism such that the images of $0$ and $1$ have the same number of $1$'s is
	running-sum synchronized.

	\begin{theorem}
		Suppose ${\bf s} = (s(n))_{n \geq 0}$ is a $k$-automatic sequence taking values in $\Sigma_2$ that is running-sum synchronized.  Then there is an automaton accepting, in parallel, 
		the base-$k$ representations of those pairs $(i,n)$
		for which ${\bf s}[i..i+n-1]$ is Dyck.  Furthermore, there is an automaton accepting, in parallel, the
		base-$k$ representations of those triples $(i,n,x)$
		for which ${\bf s}[i..i+n-1]$ is Dyck and whose
		nesting level is $x$.  In both cases, the automaton can be effectively constructed.
		\label{autothm}
	\end{theorem}
	
	\begin{proof}
		We use the fact that it suffices to create first-order logical formulas for these claims \cite{Shallit:2022}.
		Suppose $V(n,x)$ is true
		if and only $v(n) = x$.  Then define
		\begin{align*}
			N_1(i,n,x): & \  \exists y, z \ V(i,y) \andd V(i+n,z) \andd x+y = z \\
			N_0(i,n,x): &\  \exists y \ N_1(i,n,y) \andd n=x+y \\
			\Dyck(i,n): & \ (\exists w \ N_0(i,n,w) \andd N_1(i,n,w))
			\andd \\
			&(\forall t,y,z\ (t<n \andd N_0(i,t,y) \andd N_1(i,t,z)) \implies y \geq z) .
		\end{align*}
		Here 
		\begin{itemize}
			\item $N_0(i,n,x)$ asserts that $|{\bf s}[i..i+n-1]|_0 = x$;
			\item $N_1(i,n,x)$ asserts that $|{\bf s}[i..i+n-1]|_1 = x$;
			\item $\Dyck(i,n)$ asserts that ${\bf s}[i..i+n-1]$ is Dyck.
		\end{itemize}
		We can now build an automaton for $\Dyck(i,n)$ using
		the methods discussed in \cite{Shallit:2022}.
		
		Next we turn to nesting level.  
		First we need a first-order formula for the balance $B(x)$ of a factor $x$.  Since we are only interested in balance for prefixes of Dyck words, it suffices to compute $\max(0, B(x))$ for a factor $x$.    We can do this
		as follows:
		$$\Bal(i,n,x):  \ \exists y,z \ N_0(i,n,y) \andd N_1(i,n,z) \andd ((y<z \andd x=0) \mid (y\geq z \andd y=x+z)) .$$
		
		Next, we compute the nesting level of a factor, assuming it is Dyck:
		$$ \Nest(i,n,x):  \ \exists m\ m<n \andd \Bal(i,m,x) \andd \forall p,y\ (p<n \andd \Bal(i,p,y)) \implies y\leq x.$$
		This completes the proof.
	\end{proof}
	
	\begin{corollary}
		If ${\bf s} = (s(n))_{n \geq 0}$ is a $k$-automatic sequence taking values in $\Sigma_2$ that is running-sum synchronized, then it is decidable 
		\begin{itemize}
			\item[(a)] whether $\bf s$ has arbitrarily large Dyck factors;
			\item[(b)] whether Dyck factors of
			$\bf s$ are of unbounded nesting level.
		\end{itemize}
	\end{corollary}\enlargethispage{\baselineskip}
	
	\begin{proof}
		It suffices to create first-order logical statements asserting the two properties:
		\begin{itemize}
			\item[(a)] $\forall n \ \exists i,m \ m>n \andd \Dyck(i,m) $
			\item[(b)] $\forall q \ \exists i,n,p  \ 
			\Dyck(i,n) \andd \Nest(i,n,p) \andd p>q $. \qedhere
		\end{itemize}
	\end{proof}
	
	\begin{example}
		As an example, let us use {\tt Walnut} to prove that
		there is a Dyck factor of the Thue-Morse word for all even lengths.
		We can use the following {\tt Walnut} commands, which implement the ideas above.  We use the fact that the sum of $T[0..n-1]$ is $n/2$ if $n$
		is even, and $(n-1)/2 + T[n-1]$ if $n$ is odd.
		\begin{verbatim}
			def even "Ek n=2*k":
			def odd "Ek n=2*k+1":
			def V "($even(n) & 2*x=n) | ($odd(n) & 2*x+1=n & T[n-1]=@0) | 
			($odd(n) & 2*x=n+1 & T[n-1]=@1)":
			# number of 1's in prefix T[0..n-1]
			
			def N1 "Ey,z $V(i,y) & $V(i+n,z) & x+y=z":
			# number of 1's in T[i..i+n-1]
			def N0 "Ey $N1(i,n,y) & n=x+y":
			
			def Dyck "(Ew $N0(i,n,w) & $N1(i,n,w)) & 
			At,y,z (t<n & $N0(i,t,y) & $N1(i,t,z)) => y>=z":
			# is T[i..i+n-1] a Dyck word?
			
			eval AllLengths "An $even(n) => Ei $Dyck(i,n)":
		\end{verbatim}
		and {\tt Walnut} returns {\tt TRUE}.
	\end{example}
	
\begin{figure}[h]
		\centering
		\includegraphics[scale=0.6]{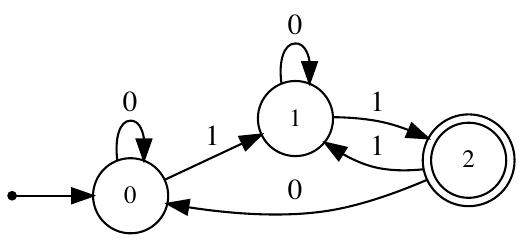}
		\caption{DFA accepting base-$2$ representations of $i$ such that the Thue-Morse word has arbitrarily long Dyck factors starting at index $i$.}
		\label{Fig:TMStartLong}
	\end{figure}

	\begin{example}
		Continuing the previous example, let us prove some other interesting statements about the Dyck factors of the Thue-Morse word. 
		
		First we show that the nesting level of every Dyck factor of Thue-Morse is $\leq 2$.  Of course, this follows from Theorem~\ref{dycktm}, but this shows how it can be done for any automatic sequence that is running-sum synchronized.  We use the following {\tt Walnut} commands:
		\begin{verbatim}
			def Bal "Ey,z $N0(i,n,y) & $N1(i,n,z) & 
			((y<z & x=0) | (y>=z & y=x+z))":
			# computes max(0, B(T[i..i+n])) where B is balance; 14 states
			def Nest "Em (m<n) & $Bal(i,m,x) & 
			Ap,y (p<n & $Bal(i,p,y)) => y<=x":
			# computes nesting level of factor, assuming it is Dyck
			
			eval maxnest2 "Ai,n,x ($Dyck(i,n) & $Nest(i,n,x)) => x<=2":
		\end{verbatim}
		and {\tt Walnut} returns {\tt TRUE} for the
		last assertion.
		
		We now consider two questions about the indices at which Dyck factors start in the Thue-Morse word.  First of all, we show that there is a Dyck word starting at every index $i$ such that $T[i]=0$.  (The condition that $T[i]=0$ is obviously necessary.)  We use the following {\tt Walnut} command:
		\begin{verbatim}
			eval everyindex "Ai T[i]=@0 => En (n>0) & $Dyck(i,n)":
		\end{verbatim}
		and {\tt Walnut} returns {\tt TRUE}.   We also describe the indices at which there are arbitrarily long Dyck factors starting in the Thue-Morse by means of an automaton.  We use the following {\tt Walnut} command:
		\begin{verbatim}
			def startlong "Am En (n>m) & $Dyck(i,n)":
		\end{verbatim}
		and {\tt Walnut} returns the $3$-state automaton in Figure~\ref{Fig:TMStartLong}, which accepts base-$2$ representations of $i$ such that the Thue-Morse word has arbitrarily long Dyck factors starting at index $i$.  In particular, we observe that there are infinitely many indices at which arbitrarily long Dyck factors start in the Thue-More word.
	\end{example}
	
	Now we turn to enumerating Dyck factors by
	length.
	Let us recall that a sequence 
	$(s(n))_{n\geq 0}$ is {\it $k$-regular\/} if
	there is a finite set of sequences
	$(s_i(n))_{n \geq 0}$, $i = 1, \ldots, t$,
	with $s = s_1$, such that
	every subsequence of the form 
	$(s(k^e n +a))_{n \geq 0}$ with $e \geq 0$
	and $0 \leq a < k^e$ can be expressed as a linear
	combination of the $s_i$.   See 
	\cite{Allouche&Shallit:1992} for more details.
	
	Alternatively, a sequence $(s(n))_{n \geq 0}$ is $k$-regular if there is
	a linear representation for it.  
	If $v$ is a
	row vector of dimension $t$, $w$ is a column
	vector of dimension $t$, and $\gamma$ is a matrix-valued morphism with domain $\Sigma_k$
	and range $t \times t$-matrices, then we say that the triple $(v, \gamma, w)$ is a {\it linear representation\/} for a function $s(n)$, of rank $t$.  It is
	defined by $s(n) = v \gamma(x) w$, where $x$ is any
	base-$k$ representation of $n$ (i.e., possibly containing leading zeroes).
	See \cite{Berstel&Reutenauer:2011} for more details.
	
	It is not difficult to use the characterization of Theorem~\ref{dycktm} to find a linear representation
	for $d(n)$, the number of Dyck factors of length $2n$
	appearing in $\bf t$, the Thue-Morse word.  However,
	in this section we will instead use a different approach that is more general.
	
	\begin{theorem}
		Suppose ${\bf s} = (s(n))_{n \geq 0}$ is a $k$-automatic sequence that is running-sum synchronized.
		Then $(d(n))_{n \geq 0}$, the number
		of Dyck factors of length $2n$ appearing in $\bf s$,
		is $k$-regular.
		\label{autocor}
	\end{theorem}
	
	\begin{proof}
		It suffices to find a linear representation for $d(n)$.  
		
		To do so, we first find a first-order formula asserting that ${\bf s}[i..i+n-1]$ is {\it novel\/}; that is, it is the first occurrence of this factor in
		$\bf s$:
		\begin{align*}
			\FacEq(i,j,n):   & \ \forall t \ (t<n) \implies {\bf s}[i+t] = {\bf s}[j+t] \\
			\Novel(i,n): & \ \forall j \ \FacEq(i,j,n) \implies j \geq i.
		\end{align*}
		Then the number of $i$ for which
		$$ \Novel(i,2n) \andd \Dyck(i,2n)$$
		holds is precisely the number of Dyck factors of $\bf s$ of length $2n$.   
		Since $\bf s$ is $k$-automatic, and its running sum sequence $\bf v$ is synchronized, it follows that there is an automaton
		recognizing those $i$ and $n$ for which
		$ \Novel(i,2n) \andd \Dyck(i,2n)$ evaluates to true, and
		from known techniques we can construct a linear
		representation for the number of such $i$.
	\end{proof}

	\begin{corollary}
		Let $d(n)$ denote the number of Dyck factors of length $2n$ appearing in the Thue-Morse word.   Then $(d(n))_{n\geq 0}$ is a $2$-regular sequence.
	\end{corollary}
	
	\begin{proof}
		We can carry out the proof of Theorem~\ref{autocor} in
		{\tt Walnut} for $\bf t$, as follows:  
		\begin{verbatim}
			def FacEq "At (t<n) => T[i+t]=T[j+t]":
			def Novel "Aj $FacEq(i,j,n) => j>=i":
			def NovelDyck "$Dyck(i,n) & $Novel(i,n)":
			def LR n "$NovelDyck(i,2*n)":
		\end{verbatim}
		The last command creates a rank-29 linear representation for the number of length-$2n$ Dyck factors. 
	\end{proof}
	
	\begin{remark}
		Using the algorithm of Sch\"utzenberger discussed
		in \cite[Chapter 2]{Berstel&Reutenauer:2011}, we
		can minimize the linear representation obtained in the proof to find a linear representation
		$(v_d, \gamma_d, w_d)$ for $d$ of
		rank $7$, as follows:
		
		\begin{align*}
			v_d^T &= \left[ \begin{smallarray}{ccccccc}
				1&0&0&0&0&0&0
			\end{smallarray} \right] 
			& \quad
			\gamma_d(0) & = \left[ \begin{smallarray}{ccccccc}
				1&    0&    0&    0&    0&    0&    0    \\
				0&    0&    1&    0&    0&    0&    0    \\
				0&    0&    0&    0&    1&    0&    0   \\
				0&    0&    0&    0&    0&    0&    1    \\
				0&    0&    0&   -2&    3&   -2&    2  \\
				0&    0&    0&    0&    2&   -2&    2   \\
				0   & 0   & 0   & 1/2 & 5/4& -5/2 & 3   
			\end{smallarray} \right] \\
			\gamma_d(1) &= \left[\begin{smallarray}{ccccccc} 
				0&     1&     0&     0&     0&     0&     0    \\
				0&     0&     0&     1&     0&     0&     0   \\
				0&     0&     0&     0&     0&     1&     0  \\
				0    & 0    & 0    & 3/4 & 11/8&-2    & 3/2  \\
				0   &  0   &  0   &  1/2 &  1/4&  0   &  1     \\
				0   &  0   &  0   & -5/2 &  11/4& -2   &  3    \\
				0   &  0   &  0   & -7/2 &  19/4& -5   &  5    \\
			\end{smallarray}\right]
			& \quad
			w_d &= \left[ \begin{smallarray}{c}
				1\\
				1\\
				2\\
				3\\
				2\\
				4\\
				6
			\end{smallarray} 
			\right] .
		\end{align*}
		This gives a very efficient way to compute $d(n)$.
	\end{remark}

	Table~\ref{tab1} gives the first few terms
	of the sequence $d(n)$.  It is sequence 
	\seqnum{A345199}
	in the {\it On-Line Encyclopedia of Integer
		Sequences} \cite{Sloane:2022}.
	\begin{table}[H]
		\begin{center}\scalebox{0.9}{
				\begin{tabular}{c|ccccccccccccccccccccc}
					$n$ & 0& 1& 2& 3& 4& 5& 6& 7& 8& 9&10&11&12&13&14&15&16&17&18&19&20\\
					\hline
					$d(n)$ & 1& 1& 2& 3& 2& 4& 6& 6& 4& 8& 8& 8&12& 9&12&13& 8&14&16&14&16
			\end{tabular}}
		\end{center}
		\caption{First few values of $d(n)$.}
		\label{tab1}
	\end{table}
	
	\section{Upper and lower bounds for \texorpdfstring{$d(n)$}{d(n)}}
	
	In this section we prove tight upper and lower bounds
	for $d(n)$, the number of Dyck factors of $\bf t$
	of length $2n$.
	
	We start with a characterization of some of the subsequences
	of $(d(n))_{n \geq 0}$.
	\begin{lemma}
		We have
		\begin{align}
			d(2n) &= 2d(n) \label{a1} \\
			d(4n+3) &= 2d(n) + d(2n+1) + q(n) \label{a2} \\
			d(8n+1) &= 2d(2n+1) + d(4n+1) - q(n) \label{a3} \\
			d(8n+5) &= 2d(n) + d(2n+1) + 2d(2n+2) \label{a4} 
		\end{align}
		for all $n \geq 3$.  Here $q(n)$ is the $2$-automatic sequence computed by the DFAO
		in Figure~\ref{fig1}.
		\begin{figure}[h]
			\begin{center}
				\vspace{-0.5in}
				\includegraphics[scale=0.7]{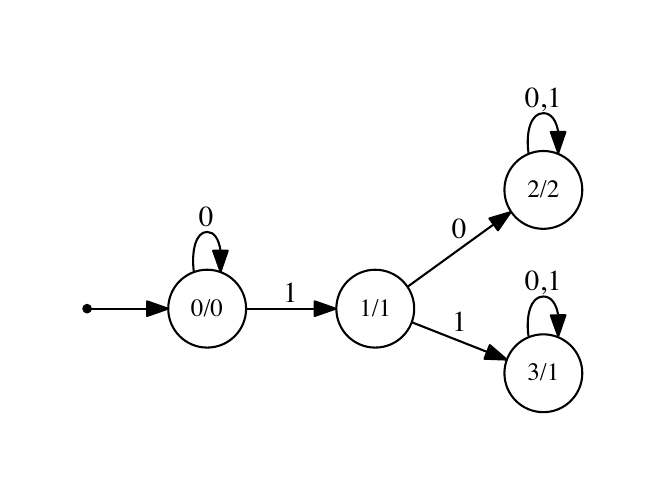}
				\vspace{-0.5in}
			\end{center}
			\caption{DFAO computing $q(n)$.  States are in the form $q/a$, where $q$ is the name of the state and $a$ is the output.}
			\label{fig1}
		\end{figure}
	\end{lemma} 
	
	\begin{proof}
		Notice that $1 \leq q(n) \leq 2$ for $n \geq 1$.
		
		These relations can be proved using linear
		representations computable by {\tt Walnut}.
		We only prove the most complicated one,
		namely Eq.~\eqref{a3}.  Substituting $n = m+3$, we see that Eq.~\eqref{a3} is equivalent to
		the claim that 
		\[d(8m+25) = 2d(2m+7) + d(4m+13) - q(m+3) \text{ for } m \geq 0.\]
		We now obtain
		linear representations
		for each of the terms, using the following
		{\tt Walnut} commands.
		\begin{verbatim}
			morphism aa "0->01 1->23 2->22 3->33":
			morphism b "0->0 1->1 2->2 3->1":
			promote Q1 aa:
			image Q b Q1:
			
			def term1 m "$LR(i,8*m+25)":
			def term2 m "$LR(i,2*m+7)":
			def term3 m "$LR(i,4*m+13)":
			def term4 m "(i=0 & Q[m+3]=@1) | (i<=1 & Q[m+3]=@2)":
		\end{verbatim}
		From these four linear representations,
		using block matrices, we can easily create a linear
		representation for 
		$$d(8m+25) - 2d(2m+7) - d(4m+13) + q(m+3).$$
		It has rank $735$.  When we minimize it (using a {\tt Maple} implementation of the Sch\"utzenberger algorithm mentioned previously), we
		get the linear representation for the $0$
		function, thus proving the identity.
		
		The other identities can be proved similarly.
	\end{proof}
	
	\begin{theorem}
		We have $d(n) \leq n$ for all $n \geq 1$.  Furthermore, this bound is tight, since
		$d(n) = n$ for $n = 3 \cdot 2^i$ and $i \geq 0$.
		\label{thm16}
	\end{theorem}
	
	\begin{proof}
		We will actually prove the stronger bound that $d(n) \leq n - (n \bmod 2)$
		for $n \geq 1$, by induction.   
		
		The base case is $1 \leq n < 29$.   In this case we can verify the bound by direct computation.
		Otherwise assume $n \geq 29$ and the bound is true for all smaller positive $n' < n$ (the $29$ comes from the fact that Eq.~\eqref{a4}
		is only valid for $n\geq 3$); we prove it for $n$.
		
		There are four cases
		to consider:  $n \equiv \modd{0} {2}$,
		$n \equiv \modd{3} {4}$, 
		$n \equiv \modd{1} {8}$,
		and $n \equiv \modd{5} {8}$.
		
		Suppose $n \equiv \modd{0} {2}$.  By induction we have
		$d(n/2) \leq n/2 - (n/2 \bmod 2)$.
		But from Eq.~\eqref{a1} we have $d(n) = 2d(n/2)
		\leq 2(n/2) - 2(n/2 \bmod 2 ) \leq n$.
		
		Suppose $n \equiv \modd{3} {4}$. By induction we have \[d({(n-3) / 4}) \leq {(n-3)/ 4} -
		({(n-3)/ 4} \bmod 2) \text{ and}\]
		\[d({(n-1)/ 2}) \leq {(n-1)/ 2} - 
		({(n-1) / 2} \bmod 2).\]   From
		Eq.~\eqref{a2} we have
		\begin{align*}
			d(n) &= 2d((n-3)/4) + d({(n-1) / 2}) + q({(n-3) / 4}) \\
			&\leq {(n-3)/2} - 2({(n-3)/4} 
			\bmod 2) + {(n-1)/2} 
			- ({ (n-1)/2} 
			\bmod 2) \\
			&\quad\quad\quad +q({(n-3) / 4}) \\
			&\leq n-1,
		\end{align*}
		as desired.
		
		Suppose $n \equiv \modd{1} {8}$.  By induction
		we have
		\[d ({(n+3)/ 4}) \leq {(n+3) / 4} - ({(n+3)/ 4} \bmod 2) \text{ and}\]
		\[d ({(n+1) / 2}) \leq {(n+1) / 2} -({(n+1)/ 2} \bmod 2).\]
		From Eq.~\eqref{a3} we have
		\begin{align*}
			d(n) &= 2 d({(n+3)/ 4}) + d({(n+1) / 2} ) - q({(n-1)/ 8}) \\
			&\leq {(n+3)/ 2} - 2({(n+3)/4} \bmod 2)
			+ {(n+1)/2} - 2({(n+1)/2} \bmod 2) \\
			&\quad\quad\quad - q({(n-1)/ 8}) \\
			&\leq n-1,
		\end{align*}
		as desired.
		
		Suppose $n \equiv \modd{5} {8}$.  By induction
		we have
		\begin{align*}
			d({(n-5)/ 8}) &\leq {(n-5)/ 8} - ({(n-5)/ 8} \bmod 2) \\
			d({(n-1)/ 4}) &\leq {(n-1)/ 4} - 
			({(n-1) / 4} \bmod 2) \\
			d({(n+3) / 4}) &\leq {(n+3)/ 4} -
			({(n+3) / 4} \bmod 2).
		\end{align*}
		From
		Eq.~\eqref{a4} we have
		\begin{align*}
			d(n) &= 2 d({(n-5)/ 8}) + d({(n-1)/ 4}) 
			+ 2d({(n+3)/ 4}) \\
			&\leq {(n-5) / 4} - 2({(n-5)/ 8} \bmod 2) + {(n-1)/ 4} - 
			({(n-1) / 4} \bmod 2) \\
			&\quad\quad\quad
			+{(n+3) / 2} - 2({(n+3) / 4} \bmod 2)
			\\
			&\leq n-1,
		\end{align*}
		as desired.
		This completes the proof of the upper bound.
		
		We can see that $d(n) = n$ for $n = 3 \cdot 2^i$
		as follows.   Using the linear representation
		for $n$ we have 
		$d(3 \cdot 2^i) = v_d \gamma_d(11) \gamma_d (0)^i w_d$.   
		
		The minimal polynomial of
		$\gamma_d (0)$ is $X^2 (X-1)(X+1)(X-2)$.
		It follows that 
		\[d(3 \cdot 2^i) = a \cdot 2^i + b + c(-1)^i \text{ for }i \geq 2.\]  Solving for the constants,
		we find that $a = 3$, $b = 0$, $c = 0$,
		and hence $d(3 \cdot 2^i) = 3 \cdot 2^i$
		as claimed.
	\end{proof}
	
	\begin{theorem}
		We have $d(n) \geq n/2$ for $n \geq 0$, and
		$d(n) \geq (n+3)/2$ for $n\geq 1$ odd.  Furthermore, the bound $d(n) \geq n/2$ is
		attained infinitely often.
	\end{theorem}
	
	\begin{proof}
		We prove the result by induction on $n$.
		It is easy to verify by direct computation
		that the result is true for $n < 29$.
		Otherwise assume $n \geq 29$ and the bound is true for all small positive $n'<n$; we prove it for $n$.
		
		Again we consider the four cases
		$n \equiv \modd{0} {2}$,
		$n \equiv \modd{3} {4}$, 
		$n \equiv \modd{1} {8}$,
		and $n \equiv \modd{5} {8}$.
		
		Suppose $n \equiv \modd{0} {2}$.    By induction and Eq.~\eqref{a1} we have
		\[d(n) = 2 d(n/2) \geq 2 (n/2)/2 = n/2.\]
		
		Otherwise $n$ is odd.
		
		Suppose $n \equiv \modd{3} {4}$.   By induction
		we have 
		\begin{align*}
			d((n-3)/4) &\geq (n-3)/8 \\
			d((n-1)/2) &\geq (n+5)/4. 
		\end{align*}
		Hence, using Eq.~\eqref{a2} we get
		\begin{align*}
			d(n) &= 2d((n-3)/4) + d((n-1)/2) + q((n-3)/4) \\
			&\geq (n-3)/4 + (n+5)/4 + q((n-3)/4) \\
			& \geq (n+1)/2 + 1 \\
			& = (n+3)/2.
		\end{align*}
		Suppose $n \equiv \modd{1} {8}$.  By induction
		we have 
		\begin{align*}
			d((n+3)/4) &\geq ((n+3)/4 + 3)/2 = (n+15)/8 \\
			d((n+1)/2) &\geq ((n+1)/2 + 3)/2 = (n+7)/4.
		\end{align*}
		Hence, using Eq.~\eqref{a3} we
		get
		\begin{align*}
			d(n) &= 2d((n+3)/4) + d((n+1)/2) - q((n-1)/8) \\
			&\geq  (n+15)/4 + (n+7)/4 - 2 \\
			& = (n+7)/2.
		\end{align*}
		Suppose $n \equiv \modd{5} {8}$.  By induction
		we have 
		\begin{align*}
			d((n-5)/8) &\geq (n-5)/16 \\
			d((n-1)/4) &\geq ((n-1)/4 + 3)/2 = (n+11)/8 \\
			d((n+3)/4) &\geq (n+3)/8.
		\end{align*}
		Hence, using Eq.~\eqref{a4} we get
		\begin{align*}
			d(n) &= 2d((n-5)/8) + d((n-1)/4) + 2d((n+3)/4) \\
			&\geq 2(n-5)/16 + (n+11)/8 + 2(n+3)/8 \\
			&= (n+3)/2.
		\end{align*}
		This completes the induction proof of both lower bounds.

		It is easy to prove, using the same techniques
		as in the last part of the proof of Theorem~\ref{thm16}, that
		$d(n) = n/2$ for $n = 2^i$, $i \geq 2$.  
	\end{proof}
	
	\begin{theorem}
		We have $\sum_{0 \leq i < 2^n} d(i) = 19\cdot 4^n/48 - 2^n/4 + 5/3$ for $n \geq 2$.
	\end{theorem}
	
	\begin{proof}
		The summation $\sum_{0 \leq i < 2^n} d(i)$
		is easily seen to equal
		$v_d (\gamma_d (0) + \gamma_d (1))^n w_d$.
		We can then apply the same techniques as above to
		the matrix $\gamma_d (0) + \gamma_d (1)$.
	\end{proof}
	
	It follows that the ``average'' value of
	$d(n)$ is $\frac{19}{24} n$.
	
	%\section{Dyck words are not first-order expressible}
	
	%In this section we give another proof, %different from that in ???, to show %that the property of being a Dyck word %is not first-order expressible.
	
	%To do so, we find a $k$-automatic %sequence such that those pairs $(i,n)$ with ${\bf x}[i..i+n-1]$ a Dyck word do %not form an automatic sequence.
	
	%??? how to do that?
	
	\section{Dyck words in other sequences}
	
	\begin{proposition}
		The only nonempty Dyck words in the Fibonacci word {\bf f} are $01$ and $0101$.
	\end{proposition}
	
	\begin{proof}
		Let $\theta$ be the Fibonacci morphism defined by $\theta(0)=01$ and $\theta(1)=0$.
		Let $w$ be a nonempty Dyck factor of the Fibonacci word.  Then $w$ begins with $0$,
		ends with $1$, and has an equal number of $0$'s and $1$'s.  It follows that $w=\theta(w')$,
		where $w'$ is a factor of the Fibonacci word consisting entirely of $0$'s.
		However, the longest such $w'$ is $w'=00$.
	\end{proof}
	
	A similar argument applied to the morphism that maps $0 \to 01$ and $1 \to 00$ gives
	the following result.
	
	\begin{proposition}
		The only nonempty Dyck words in the period-doubling sequence are 
		$01$, 
		$0101$, and 
		$010101$.
	\end{proposition}
	
	Recall that the Rudin-Shapiro sequence 
	${\bf r}= (r(n))_{n \geq 0}$
	is defined to be the number
	of occurrences of $11$, taken modulo $2$, in
	the base-$2$ expansion of $n$.  
	
	\begin{theorem}
		There are Dyck factors of arbitrarily large nesting level in the Rudin-Shapiro sequence.
	\end{theorem}
	
	\begin{proof}
		For $n\geq 0$ define $x_n = {\bf r}[2\cdot 4^n..4^{n+1}-1]$.
		We will show, by induction on $n$, that
		$x_n$ is a Dyck factor of nesting level
		$2^{n+1} - 1$.
		
		The base case is $n=0$.  In this case ${\bf r}[2..3] = 01$ is a Dyck factor of nesting level $1$.
		
		%Now assume the result is true for $n$; we will prove it for $n+1$.  
		For $n \geq 0$ define
		$y_n = {\bf r}[0..2\cdot 4^n - 1]$.   We claim that
		$x_{n+1} = y_n x_n \overline{y_n} x_n$; this follows
		immediately by considering the first three bits
		of the base-$2$ representations of the numbers
		in the range $[2 \cdot 4^{n+1} .. 4^{n+2}-1]$.
		
		Define $s(n) = \sum_{0 \leq i \leq n} (-1)^{r(i)}$.
		It should be clear that $s(n)$ is the imbalance between the number of $0$'s (left parens) and $1$'s (right parens) in ${\bf r}[0..n]$.
		We now claim that $0 < s(i) \leq s(2 \cdot 4^n - 1) = 2^{n+1}$ for $0 \leq i \leq 2 \cdot 4^n - 1$.  In fact, the stronger claim $s(i) > 0$ for all $i$
		is \cite[Satz 9]{Brillhart&Morton:1978}. The fact that $s(2 \cdot 4^n - 1) = 2^{n+1}$ is
		\cite[Beispiel 6]{Brillhart&Morton:1978}, and
		the inequality $s(i) \leq 2^{n+1}$ for $0 \leq i \leq 2 \cdot 4^n - 1$ can be deduced from
		\cite[Satz 9]{Brillhart&Morton:1978}.  Thus we have shown that the imbalance of $y_n$ is $2^{n+1}$,
		the imbalance of $x_n$ is $0$ and its nesting
		level is $2^{n+1} - 1$, the imbalance of $\overline{y_n}$ is $-2^{n+1}$, and hence
		$x_{n+1} = y_n x_n \overline{y_n} x_n$ is Dyck
		with nesting level $2^{n+2} - 1$.
	\end{proof}

	\begin{theorem}
		The set of $n$ such that there is a Dyck factor of length $n$ in the Rudin-Shapiro word is a $4$-automatic (and hence $2$-automatic) set.
	\end{theorem}
	
	\begin{proof}
		Our proof uses Walnut.  However, the reader will recall from our earlier discussion
		that we need ${\bf r}$ to be running-sum synchronized (i.e., there is an
		automaton accepting in parallel the base-$2$ representations of
		$n$ and $\sum_{0\leq i < n}r(n)$) in order for us to be able to apply Walnut.
		It turns out that ${\bf r}$ is not running-sum synchronized for base $2$.
		However, in \cite{Rampersad&Shallit:2023} the last two
		authors show that ${\bf r}$ is $(4,2)$-running-sum synchronized; i.e., there is an
		automaton accepting in parallel the representations of $n$ and $\sum_{0\leq j \leq n}r(j)$,
		where $n$ is given in base-$4$ and the running sum\footnote{The running sum here is
			somewhat unusually indexed as running from $0$ to $n$ rather than from $0$ to $n-1$, which leads
			to the awkward appearance of various $+1$ or $-1$ terms in our Walnut formula.}
		is given in base-$2$.
		
		\begin{figure}[h]
			\begin{center}
				\vspace{-0.3in}
				\includegraphics[scale=0.4]{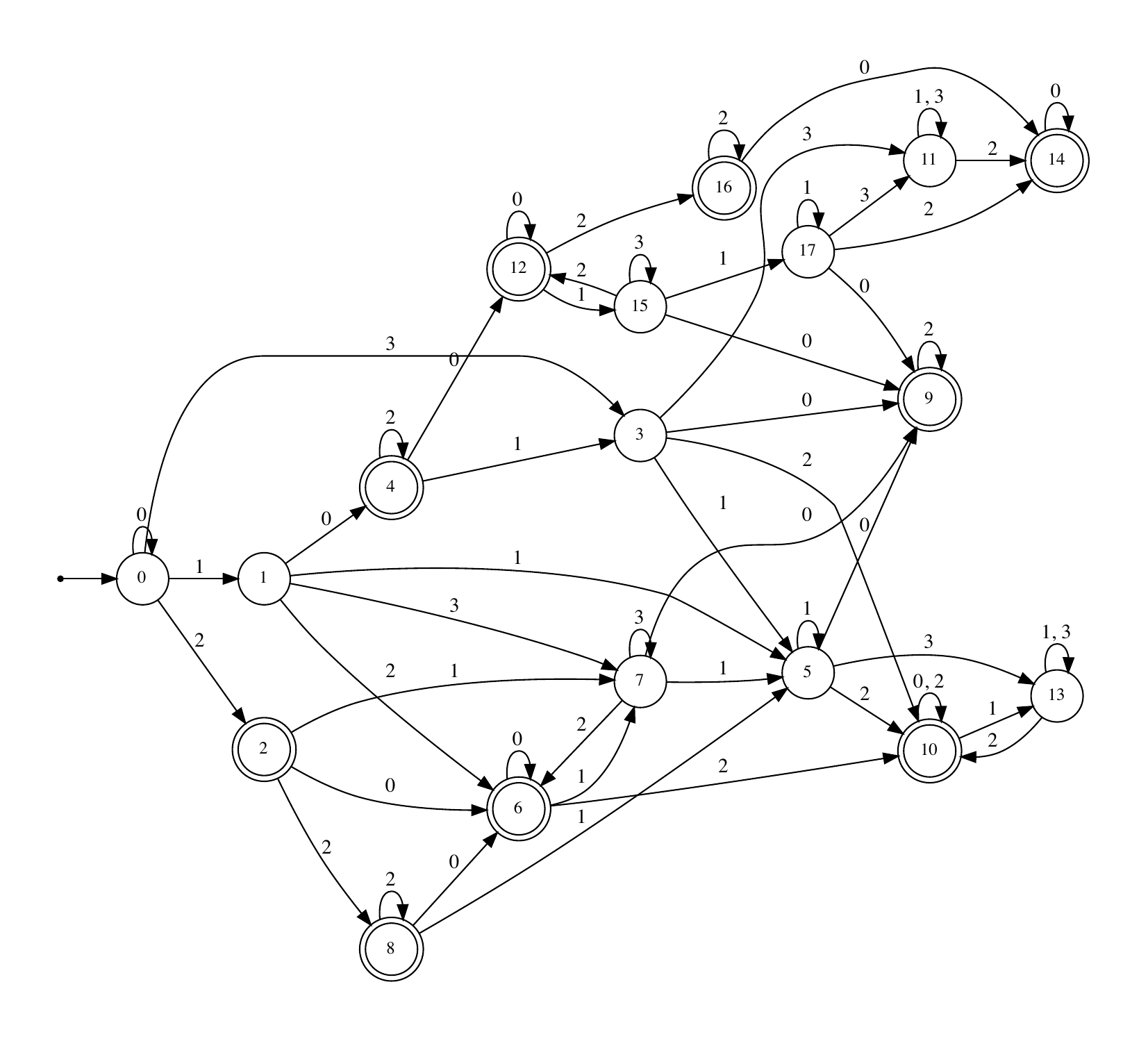}
				\vspace{-0.4in}
			\end{center}
			\caption{DFA accepting base-$4$ representations of $n$ such that the Rudin--Shapiro sequence contains a Dyck factor of length $n$.}
			\label{fig:dyck_rs}
		\end{figure}
		The Walnut code given below computes an automaton (Figure~\ref{fig:dyck_rs})
		accepting the base-$4$ representations
		of all $n$ such that there is a Dyck factor of length $n$ in the Rudin-Shapiro word.
		Here \texttt{RS4} refers to a DFAO that takes the base-$4$ representation of $n$ as input
		and computes the $n$-th term of the Rudin-Shapiro sequence over $\{+1,-1\}$ (so here $+1$ plays
		the role of the left parenthesis and $-1$ plays the role of the right parenthesis).
		The command $\$\mathtt{rss(i,x)}$ refers to an invocation of the automaton given
		in \cite{Rampersad&Shallit:2023} for the running sum function; i.e., this command returns \texttt{TRUE}
		if $x$ is the base-$2$ representation of $\sum_{0\leq j \leq i}r(j)$ and $i$ is given
		in base-$4$.
		
		\enlargethispage{\baselineskip}
		\begin{verbatim}
			eval dyck_rs "Ei ?msd_4 n>=1 &
			(Ax,y ($rss(i,x) & $rss(i+n-1,y) & RS4[i] = @1) =>
			?msd_2 x=y+1) &
			(Ax,y ($rss(i,x) & $rss(i+n-1,y) & RS4[i] = @-1) =>
			?msd_2 x=y-1) &
			(Ax,y,t  (t<n & $rss(i,x) & $rss(i+t,y) & RS4[i] = @1) =>
			?msd_2 x<=y+1) &
			(Ax,y,t  (t<n & $rss(i,x) & $rss(i+t,y) & RS4[i] = @-1) =>
			?msd_2 x<=y-1)":
		\end{verbatim}
	\end{proof}
	
	We also offer the following conjecture concerning the Dyck factors of the paperfolding
	sequence.
	
	\begin{conjecture}
		The paperfolding sequence has a Dyck factor of length $n$ iff $n$ is of the form $2^k - 2^i$
		for $0 \leq i < k$.
	\end{conjecture}
	
	\subsection*{Acknowledgments}
	Research of Lucas Mol is supported by an NSERC Grant, Grant number RGPIN-2021-04084.
	Research of Narad Rampersad is supported by an NSERC Grant, Grant number 2019-04111.
	Research of Jeffrey Shallit is supported by an NSERC Grant, Grant number 2018-04118.
	
	%%% REFERENCES %%%
	%{\small\bibliography{dyck_factors}}

	\EditInfo{December 15, 2023}{May 27, 2024}{Rigo Michel, Emilie Charlier and Julien Leroy}
	
\end{document}